\documentclass{desyproc}

\begin{document}
%------------------------------------
\title{Evidence of vacuum birefringence from the polarisation of the optical emission from an Isolated Neutron Star}

%for single authors the superscripts are optional
\author{{\slshape Roberto P. Mignani$^{1,2}$, Vincenzo Testa$^{3}$, Denis Gonz\'alez Caniulef$^{4}$,  Roberto Taverna$^{5}$, Roberto Turolla$^{5,4}$,  Silvia Zane$^{4}$, Kinwah Wu$^{4}$, Gaspare Lo Curto$^{6}$}\\[1ex]
$^{1}$ INAF - Istituto di Astrofisica Spaziale e Fisica Cosmica Milano, Milano, Italy\\
$^{2}$ Janusz Gil Institute of Astronomy, University of Zielona G\'ora, Zielona G\'ora, Poland \\
$^{3}$ INAF - Osservatorio Astronomico di Roma,  Monteporzio, Italy \\
$^{4}$ Mullard Space Science Laboratory, University College London, Holmbury St. Mary, UK \\
$^{5}$ Dipartimento di Fisica e Astronomia, Universit\'a di Padova, Padova, Italy \\
$^{6}$ European Southern Observatory, Santiago de Chile, Chile
}

% if the proceedings are available online (e.g. at Indico)
% please enter the contribution ID or file_name below for the DOI
%\contribID{32}
\contribID{Mignani\_Roberto}

% TO THE CONFERENCE EDITORS: 
% please update the following information      
% before sending the template to the authors
\confID{13889}  % if the conference is on Indico uncomment this line
\desyproc{DESY-PROC-2017-XX}
\acronym{Patras 2017} % if you want the Acronym in the page footer uncomment this line
\doi  % if there is an online version we will register DOIs

\maketitle

\begin{abstract}
Isolated Neutron Stars are some of the most exciting stellar objects known to astronomers: they have the most extreme magnetic fields, with values up to $10^{15}$ G, and, with the exception of stellar-mass black holes, they are the most dense stars, with densities of $\approx 10^{14}$ g cm$^{-3}$.  As such, they are perfect laboratories to test theories of electromagnetism and nuclear physics under conditions of magnetic field and density unattainable on Earth. In particular, the interaction of radiation with strong magnetic fields is the cause of the {\em vacuum birefringence}, an effect predicted by quantum electrodynamics in 1936 but that lacked an observational evidence until now. Here, we show how the study of the polarisation of the optical radiation from the surface of an isolated neutron star yielded such an observational evidence, opening exciting perspectives for similar studies at other wavelengths.
\end{abstract}

\section{Introduction to isolated neutron stars}

Isolated neutron stars are stellar corpses left over after supernova explosions of stars about ten times as massive as our Sun.  As a consequence of the explosion, the star external layers are ejected into space to form a supernova remnant and the core of the star, $\sim$1.5 Sun masses, collapses into a sphere of about 10 km radius, of density $\sim 10^{14}$ g cm$^{-3}$, comparable to the atomic nucleus, under which free protons and electrons merge to form neutrons via inverse $\beta$ decay. 
%For the conservation of the angular momentum, 
The core is spun up to periods of a few tens of milliseconds and
%, for the conservation of the magnetic field density, 
the star magnetic field is amplified up to $10^{15}$ G, thousands billion times stronger than the Earth's. The  formation of  an isolated neutron star (INS) from a supernova explosion  was predicted in the 1930s by astronomers Walter Baade and Fritz Zwicky, soon after the discovery of the neutron by physicist James Chadwick and, possibly, following an original intuition by physicist Lev Landau.\footnote{A very nice and vivid review of the true story of how the neutron star idea developed is given in Yakovlev et al.\ 2013, Physics Uspekhi, 56, 3, arXiv:1210.0682.} 

The first INS was discovered  50 years ago \cite{hew68} as a source of pulsed radio emission,
dubbed {\em pulsar}, whose period (1.33 s) could only be explained by the fast rotation of a star with a density exactly as expected for an INS. The discovery of the Crab and Vela pulsars in their associated supernova remnants closed the loop and proved that INSs indeed form after supernova explosions.
%, as predicted by Baade and Zwicky. 
The measurement of a tiny but regular decrease of the spin period for a few pulsars suggested that their rotational energy decreases with time. Pulsar properties are generally described by the {\em magnetic dipole model} \cite{gold68,pac68}, where the radio emission is powered by the acceleration of particles at the expenses of the neutron star rotational energy, producing beamed radiation along its magnetic field axis which is seen as an intermittent signal as the beam crosses the line of sight (LOS) to the Earth if the neutron star rotation and magnetic axis are misaligned. From the magnetic dipole model, by equating the neutron star rotational energy loss rate to the magnetic dipole loss one can infer (under certain assumptions on the neutron star mass and radius) the value of the neutron star magnetic field $B_{\rm s} = 3.2 \times 10^{19} (P_{\rm s} \dot{P}_{\rm s})^{1/2}$ G, where $P_{\rm s}$ and $\dot{P}_{\rm s}$ are its spin period and first derivative, respectively. 

Pulsars feature a very complex multi-wavelength phenomenology spanning from the radio band, where they were discovered \cite{hew68} and traditionally observed, to the optical band, to the X-rays, up to the very high-energy $\gamma$-rays \cite{2pc}. From one hand, this made it possible to study the emission from the neutron star magnetosphere across the entire electromagnetic spectrum, offering a broad view on the emission mechanisms therein. On the other hand, this showed the existence of pulsars that do not emit in radio, hence dubbed {\em radio-silent}.  Multi-wavelength observations also led to the discovery of classes of INSs that are not powered by the rotational energy\cite{har13} but by, e.g. the magnetic energy, such as the {\em magnetars}\cite{mer17}, or the release of thermal radiation from the hot ($T_{\rm s} \sim 10^{5}$--$10^{6}$ K) and cooling neutron star surface. Seven of such {\em cooling} INSs are known, detected both in the soft X-rays (0.05--0.1 keV) and in the optical/ultraviolet. They have long spin periods ($P_{\rm s}$ = 3--11 s) compared to most INSs and they are endowed with magnetic fields of $10^{13}$--$10^{14}$ G  \cite{turolla09}.

\section{Scientific motivation}

The study of the thermal radiation from these INSs is the only way to peek directly at (or close to) the star surface. Indeed, we do not know yet whether the thermal emission is produced by the bare star surface or it is mediated by a thin atmosphere, and what the composition of such an atmosphere would be. Measurements of the polarisation degree of the thermal radiation from the star surface can help to address these issues \cite{denis16}, as well as to test quantum electrodynamic (QED) effects close to the neutron star surface, such as {\em vacuum birefringence} \cite{heisen36}. In brief, electromagnetic radiation in vacuum propagates along two modes, the ordinary mode (O mode), where the electric field oscillates parallel to the magnetic field plane, and the extraordinary mode (X-mode), where the electric field oscillates perpendicular to the magnetic field plane. The presence of a strong magnetic field, however, induces the formation of virtual electron/positron pairs, which change the refraction indices along the X and O modes, proportionally to the square of the magnetic field strength, and affects their propagation, changing the polarisation degree of the electromagnetic radiation.  In ground-based laboratories magnetic fields B of $\sim 10^{6}$ G at most can be generated, and vacuum birefringence effects, when they are eventually measured, will only be tested in the weak field regime  \cite{dv16}. On the other hand, close to the INS surface these effects can be tested in the strong field regime. Indeed, vacuum birefringence is expected to increase dramatically the linear polarisation degree of the thermal radiation produced from the star surface \cite{pot14}, from a level of a few per cent, up to 100 per cent, depending on the viewing geometry and the emission mechanisms \cite{heyl00,heyl02,heyl03}. The best target among the seven INSs with purely thermal emission is RX\, J1856.5$-$3754. This source has a magnetic field of $\sim 10^{13}$ G, is quite bright in the X-rays, but rather faint in the optical (magnitude 25.5, about 100 million times fainter than the naked eye limit), even if it is still the brightest of the seven. Although polarisation measurements in the X-rays should be the obvious choice for this experiment, no dedicated X-ray polarimetry satellite is currently operational, with the first one, the NASA's {\em Imaging X-ray Polarimetry Explorer} \cite{weiss16} to be launched early next decade with possibly the ESA's  {\em X-ray Imaging Polarimeter Explorer} \cite{sof16} and the chinese {\em enhanced X-ray Timing Polarimetry} satellite \cite{zhang16}, due to follow. However, none of them will be sensitive in the soft X-ray band. Therefore, we opted for a measurement of the linear polarisation in the optical, exploiting a mature technique in optical astronomy.

%As such, they are formidable laboratories where physics can be studied at its most extreme, with implications  beyond astrophysics (e.g., Mignani et al.\ 2017). 

\section{Observations and Results}
Owing to the target faintness, we observed it with the Very Large Telescope of the European Southern Observatory, an array of four telescopes of 8.2m in diameter located at the Cerro Paranal Observatory (Chile). We used the second Focal Reducer and low dispersion Spectrograph \cite{app98} mounted  at one the VLT unit telescopes. The camera is  equipped with polarisation optics to measure linear polarisation through a Wollaston prism acting as a beam splitting analyser and two super-achromatic phase  retarder 3$\times$3 plate  mosaics installed on rotatable mountings to be moved in and out of the light path. We used four half-wave retarder plate angles of $0^\circ$, $22.5^\circ$, $45^\circ$, and $67.5^\circ$, which correspond to the retarder plate orientations relative  to the Wollaston prism. A filter with central wavelength $\lambda=555.0$ nm and width $\Delta\lambda=61.6$ nm was inserted along the light path.   We acquired a total exposure of 7920 s per retarder plate angle. 
The degree of linear polarisation of a source is calculated from the normalised Stokes parameters $P_{\mathrm U} \equiv U/I$ and $P_{\mathrm Q} \equiv Q/I$ computed from the source fluxes in the ordinary and extraordinary beams \cite{mig17} as $\mathrm{P.D.} = 
%\sqrt{
( {P_\mathrm Q}^{2} + {P_\mathrm U}^{2})^{1/2}$.
%}}$. 
%
In this way, we measured a $\mathrm{P.D.}=16.43\% \pm5.26\%$. Owing to the target faintness, we carefully verified that our measurement is not dominated by systematic errors and/or observational biases.  We tested the absolute  accuracy of our polarisation measurement against polarised calibration targets and found that 
this is accurate to $0.13\% \pm0.06\%$. In a similar way, we estimated the spurious polarisation of the polarisation optics to be $0.09\% \pm0.06\%$ from the observations of un-polarised calibration targets. Since our observations were taken in New Moon, the contamination of the sky background polarisation is also close to zero. Since our target is at a distance of 400 light years, we verified that our polarisation measurement was not affected by dust grains along the line of sight by comparing the measured $\mathrm{P.D.}=16.43\% \pm5.26\%$ with that of 42 stars in the field and found that for the latter it is consistent with zero.

\begin{figure}
%[hb]
\centerline{\includegraphics[width=1.1\textwidth]{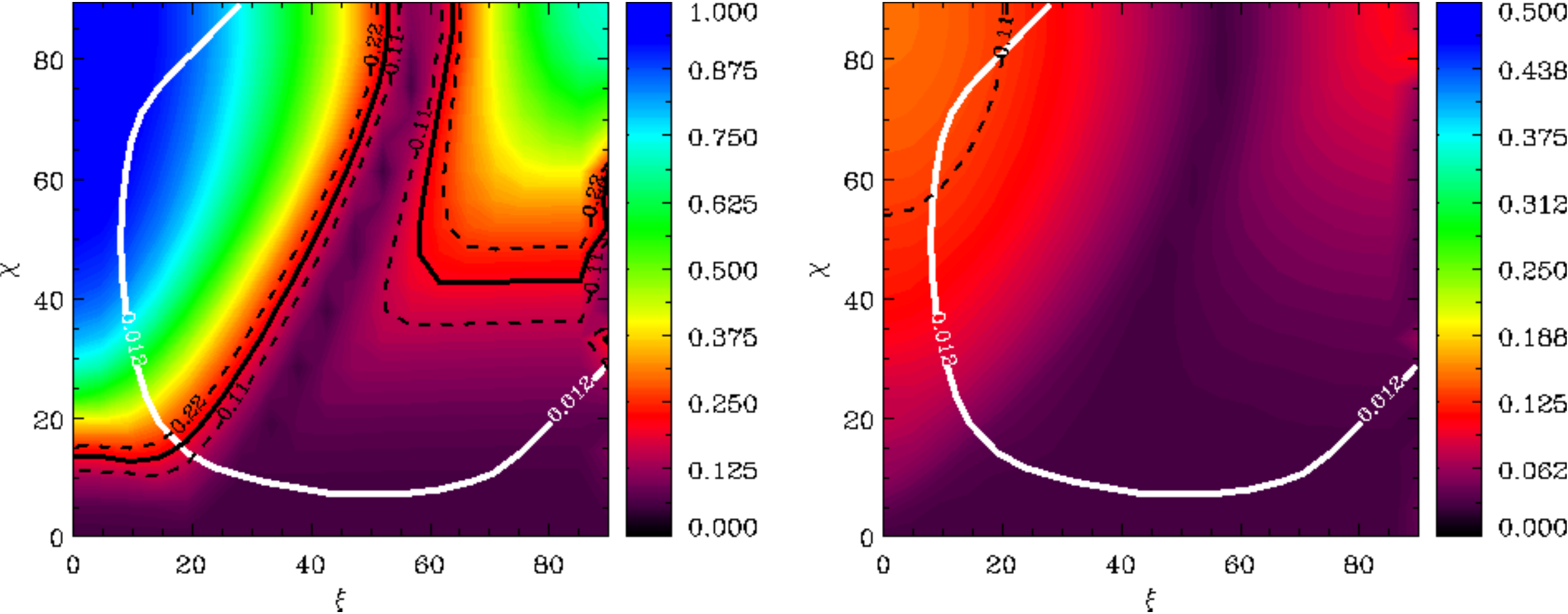}}
\caption{Simulated  $\mathrm{P.D.}$ of RX\, J1856.5$-$3754 as a function of the angles $\chi$ and $\xi$, i.e.  between the LOS and the INS spin axis and between the  INS magnetic and spin axis, respectively. A BB emission model is assumed. The white solid line represents the constraints on $\chi$ and $\xi$ inferred from the pulsed X-ray light curve  \cite{tienmer07}, whereas the black solid line corresponds to our measurement, including $1\sigma$ errors (black dashed line).  The plots on the left and on the right represent the cases when we accounted or not for QED effects, respectively.  }\label{on_off}
\label{sec:figures}
\end{figure}

%\section{Discussion}

We  compared our measurement with the predictions for four different emission models: a plain back body (BB),  a magnetised, completely ionised hydrogen atmosphere, and, a condensed surface model (both in the fixed and free ion limit), based on numerical simulations \cite{taverna15,denis16}. Figure \ref{on_off} shows the simulation corresponding to the BB case, where the $\mathrm{P.D.}$ is shown as a function of the  angles $\chi$ and $\xi$ between the LOS and the INS spin axis and between the  INS magnetic and spin axis, respectively. The plot on the left accounts for QED effects. As it can be seen, the measured $\mathrm{P.D.}$  (black solid and dashed lines) is consistent with the model predictions and the constraints on $\chi$ and $\xi$  imposed by the pulsed X-ray light curve profile \cite{tienmer07}. This is also the case for the other simulations performed for the other tested emission models \cite{mig17}. Therefore, we cannot discriminate, based on the current polarisation measurement, which of these models better describes the optical emission from RX\, J1856.5$-$3754. Decreasing the error on $\mathrm{P.D.}$ through deeper observations and deriving more constraining limits on  $\chi$ and $\xi$ is what we need to discriminate between different models. Interestingly, however,  our simulations shows that when not accounting for QED effects all of the considered emission models predicts values of $\mathrm{P.D.}$ that are only marginally consistent with our observations, see Fig. \ref{on_off} (right) for the BB case. This is a clear indication that QED effects play a crucial role in setting the actual value of the polarisation degree, see also discussion in \cite{turolla17}. Therefore, this is the first, indirect though, observational evidence of the phenomenon of {\em vacuum birefringence}, as predicted by QED over 80 years ago \cite{heisen36}. 

Follow-up polarimetry observations of RX\, J1856.5$-$3754 recently obtained at the VLT with the same instrument set-up but for a twice as long integration time will consolidate our result.
% above the $5 \sigma$ level. 
Polarimetry measurements in other filters will also be important to test the dependence of the effect on the wavelength.
%, although this might be observationally tricky. 
Measuring polarisation for other INSs similar to RX\, J1856.5$-$3754 would be important to verify the effect on sources with different magnetic field values. This is observationally challenging, however, since all these other sources are fainter than  RX\, J1856.5$-$3754, which requires a more massive investment of observing time. In the X-rays, polarimetry measurements with the upcoming X-ray polarimetry missions, sensitive between 2 and 8 keV, will make it possible to search vacuum birefringence effects in magnetars, which have an X-ray spectrum harder than RX\, J1856.5$-$3754 and even higher magnetic field values, and for which polarisation measurements in the optical are complicated by their larger distances, uncertain counterparts, and foreground polarisation contamination. 
 
\section{Acknowledgments}

RPM thanks the organisers for the invitation to participate to the PATRAS 2017 conference and acknowledges financial support from an "Occhialini Fellowship".

%\section{Bibliography}

%If possible please use the bibtex information as given by SPIRES
%to make the citations~\cite{Bertolucci:2017vgz} uniform and follow the 
%examples~\cite{Bertolucci:2017vgz,Ehret:2010mh} given below.
%Note that there is a (non-breaking) space before \verb?\cite?.

% ****************************************************************************
% BIBLIOGRAPHY AREA
% ****************************************************************************

\begin{footnotesize}

\end{footnotesize}

% ****************************************************************************
% END OF BIBLIOGRAPHY AREA
% ****************************************************************************

\end{document}